\documentclass[11pt,a4paper]{article}
\usepackage[utf8]{inputenc}
\usepackage[english]{babel}
\usepackage[T1]{fontenc}

\usepackage{newtxtext,newtxmath} %replaces mathptmx

\usepackage[pdftex]{graphicx} % Required for including pictures
\usepackage[pdftex,linkcolor=black,pdfborder={0 0 0}]{hyperref} % Format links for pdf
\usepackage{calc} % To reset the counter in the document after title page
\usepackage{enumitem} % Includes lists

\usepackage[a4paper, lmargin=0.1666\paperwidth, rmargin=0.1666\paperwidth, tmargin=0.1111\paperheight, bmargin=0.1111\paperheight]{geometry} %margins

\usepackage{parskip}
\usepackage[all]{nowidow} % Tries to remove widows
\usepackage[protrusion=true,expansion=true]{microtype} % Improves typography, load after fontpackage is selected
\usepackage{lineno}         % Line numbers
\usepackage{booktabs}       % professional-quality tables
\usepackage{amsfonts}       % blackboard math symbols
\usepackage{amsmath}    
\usepackage{nicefrac}       % compact symbols for 1/2, etc.
\usepackage{microtype}      % microtypography
\usepackage{csquotes}
\usepackage{tikz}           %tikz graphics
\usepackage{float}          %float package for figure placement
\usepackage{sidecap}        %Captions beside figures
\sidecaptionvpos{figure}{t} %Vertical alignment of side captions

\usepackage{authblk}
\usepackage{cancel}         %strikethrough symbols

\usepackage[]{natbib}

\hypersetup{ 	
pdfsubject = {},
pdftitle = {An Ansatz for computational undecidability in RNA automata},
pdfauthor = {Adam J. Svahn and Mikhail Prokopenko}
}

\title{An Ansatz for computational undecidability in RNA automata}
\author[1,2,*]{Adam J. Svahn} 
\author[1,3]{Mikhail Prokopenko}
\affil[1]{
  Centre for Complex Systems\\
  Faculty of Engineering}
\affil[2]{Westmead Clinical School\\
    Faculty of Medicine and Health} 
\affil[3]{Marie Bashir Centre for Infectious Diseases and Biosecurity}
\affil[ ]{The University of Sydney\\
Sydney\\Australia}
\affil[*]{Corresponding author: adam.svahn@sydney.edu.au}

\date{}

\begin{document} 

\maketitle

%\linenumbers

\begin{abstract}
    In this Ansatz we consider theoretical constructions of RNA polymers into automata, a form of computational structure. The basis for transitions in our automata are plausible RNA enzymes that may perform ligation or cleavage. Limited to these operations, we construct RNA automata of increasing complexity; from the Finite Automaton (RNA-FA) to the Turing Machine equivalent 2-stack PDA (RNA-2PDA) and the universal RNA-UPDA. For each automaton we show how the enzymatic reactions match the logical operations of the RNA automaton. A critical theme of the Ansatz is the self-reference in RNA automata configurations which exploits the program-data duality but results in computational undecidability. We describe how computational undecidability is exemplified in the self-referential Liar paradox that places a boundary on a logical system, and by construction, any RNA automata. We argue that an expansion of the evolutionary space for RNA-2PDA automata can be interpreted as a hierarchical resolution of computational undecidability by a meta-system (akin to Turing’s oracle), in a continual process analogous to Turing’s ordinal logics and Post’s extensible recursively generated logics. On this basis, we put forward the hypothesis that the resolution of undecidable configurations in RNA automata represent a novelty generation mechanism and propose avenues for future investigation of biological automata.
\end{abstract}

\section{Introduction}

Undecidability is an important concept in the theory of computation, where, for certain problems it can be shown that an algorithm always generating a definitive answer is impossible to construct \citep{turing_computable_1937, godel_uber_1931}. In other words, it is impossible to always decide that a given computation halts or runs forever. Undecidable dynamics, generated by self-referential relationships, have also been implicated  in chaos theory and complexity science~\citep{prokopenko_self-referential_2019}.
In this work, we pose the question of whether undecidability can be demonstrated for biological systems. 

Specifically, in this Ansatz we investigate the computational properties of RNA-based systems. With the minimal RNA-mediated functions of ligation and cleavage \citep{robertson_highly_2014, will_spliceosome_2011,evans_rnase_2006, jarrous_roles_2017} we construct theoretical RNA automata with equivalence to a finite automaton and to push-down automata with one or two stacks, with demonstration of computations achievable within each construction. Importantly, in an RNA automaton, the ribozymes that constitute the transition rules (the program) and the polymers that serve as symbolic memory (data) are both composed of the same nucleotide substrate. We describe how, from this shared substrate, an RNA based Universal Push-Down Automaton (RNA-UPDA) with equivalence to a Universal Turing Machine (UTM) may simulate any encoded RNA automaton program. 

Our main technical objective is to describe an RNA based computational framework that enables an encoding and decoding relationship that will facilitate the emergence of self-reference. It is useful to distinguish self-replication and self-reference as distinct concepts. For example, remarkable self-replicating mineral crystals which propagate patterns of inhomogeneities from layer to layer and reproduce by random fragmentation \citep{cairns-smith_origin_1966, schulman_robust_2012} are not self-referential, because the decoding relationship itself is not represented in an encoded form \citep{mcmullin_von_2012}.  On the other hand, computer programs with self-replicating code (e.g., Cellular automata) can be fully self-referential by employing \emph{explicit} encoding/decoding mechanisms. Self-reference, unlike self-replication, generates a form of undecidability exemplified by the Liar paradox wherein a self-negating statement is unsolvable within the bounds of the system. In computability theory this is manifested as the halting problem and implicated in studies of novelty generation and open-ended evolution \citep{markose_novelty_2004, kauffman_humanity_2016, markose_complex_2017, zenil_limits_2016, abrahao_algorithmic_2019, adams_formal_2017, prokopenko_self-referential_2019}. The relation between self-reference and recursive self-representation in a biological context is emphasised by Hofstadter~\citeyearpar{hofstadter_go_1980}, who highlighted that a biological self-referential system cannot be consistent. In this work we present a series of biologically plausible classes of RNA automata which reach the level at which such inconsistency and the corresponding computational undecidability is ultimately manifested. We then argue that this undecidability is framed only within specific bounds, i.e., within the corresponding formal system, and can be resolved by extending the bounds as a result of interactions between the organism and its environment, that is, as an evolutionary novelty generated by these interactions.

Goldenfeld and Woese \citeyearpar{goldenfeld_life_2011} in particular focused on self-reference to drive at the question of biological innovation :\begin{displayquote} ``...what is to us the central aspect of evolution: It is a process that continually expands the space in which it operates through a dynamic that is essentially self-referential. Self-reference should be an integral part of a proper understanding of evolution, but it is rarely considered explicitly.''\end{displayquote}

We ultimately hypothesise that an RNA automaton utilising two stacks would be capable of self-reference and so, would lead to the generation of an auto-negating undecidable `Liar paradox', recognising and resolving which would then allow the system to expand its boundaries. We hope to contribute to the early, but already productive investigations of computational principles in biological systems (reviewed in \citealp*{prohaska_how_2019}), such as the `chromatin computer' \citep{arnold_chromatin_2013}. The computational approach helps us understand extant life but also to look back at the origin of life, in particular the investigation of the evolution of evolvability \citep{virgo_lineage_2017}. The methods found at this interface of synthetic biology and simulated artificial life represent a promising test-bed for construction of computational and \emph{in-vitro} models that create self-referential, and paradoxical, scenarios from which the system must `jump out' and break the paradox by abstraction, or meta-level, simulation.

\section{Background}

An important aim of the Ansatz is to develop constructions which were plausible, meaning that the enzymes and reactions required should be already known to exist within the chemistry of single-stranded RNA molecules. In keeping with this constraint, we designed each RNA automaton to include a collection of ligation and cleavage enzymes which together form a system of reaction profiles. RNA based ligation and cleavage are well demonstrated, and some background for this is given below. We will also give a brief background of automata theory, definitions of RNA automata components and a further discussion of physical assumptions for the RNA automata.

\subsection{RNA ligation and cleavage enzymes}

The ribonucleotide monomer is a modular construction of a ribose sugar, a phosphate and a nucleobase. The nucleobase is the information carrying unit of the ribonucleotide. The nucleobases are divided into the purines (adenine and guanine) and the pyrimidines (cytosine and uracil) which form pairwise affinity relationships by hydrogen bonding; adenine to uracil and guanine to cytosine. Importantly, RNA polymers are structurally labile, readily forming complex tertiary structures. These structures create chemical micro-domains that allow the RNA polymer to act as an enzyme, facilitating a chemical reaction. RNA ligases (bond forming) and RNases (bond breaking) make use of the hydrolysing aqueous environment to catalyse the forming or the breaking of the phosphodiester bonds which link the ribonucleotide monomers together.

Ligation is a catalysed reaction that forms a bond between RNA polymers, being the joining of two polymers linearly, and is an essential function in all extant life. Remarkably, novel ligase ribozymes can be generated by \emph{in vitro} evolution \citep{joyce_protocells_2018, ekland_structurally_1995}. The R3C ligase was evolved out of a library of $10^{14}$ short RNA polymers \citep{rogers_effect_2001} and constituted a 74nt RNA polymer enzyme which ligated a target RNA polymer to itself. Importantly, this study demonstrated that a shorter R3C motif of 57nt containing the catalytic site could ligate together two opposing RNA polymers and release the product. This property was exploited to demonstrate that a redesigned R3C ligase could ligate split copies of itself \citep{paul_self-replicating_2002}, starting an auto-catalytic replication cycle \citep{paul_self-replicating_2002,lincoln_self-sustained_2009}.

We may then ask what functional role could ligation perform in a computational RNA system? The product of any ligation is the generation of an RNA polymer which is longer than the components with which the reaction began. Ribozymes perform catalytic roles through the formation of secondary and tertiary structures and the ribozyme formed from a long polymer may possess a greater propensity to form more complex and stable structures than that of a short polymer. The sequence of RNA nucleotides may also encode information to represent previous visited states of the system, with longer polymers having the potential to encode a longer sequence of symbols. The modular nature of RNA means that ligations may explore a large combinatorial space, limited by the available polymer reactants and the binding properties of the available ligating ribozymes. In combination, it can be hypothesised that ligase reaction cascades may be capable of constructing new ribozymes of increasing complexity, as well as encoding and extending symbolic representations within the system. 

If encoding is performed by ligation, then decoding is sub-served by cleavage, the splitting apart of an RNA polymer. All kingdoms of life retain a core RNA cleavage enzyme in RNase P. RNase P is a ribonucleoprotein that cleaves opposing single-stranded RNA and may bind and cleave multiple targets without losing function \citep{reiter_structure_2010}. Structural and functional study of RNase P has resulted in a consensus that the RNA component was likely to have been present in the Last Universal Common Ancestor, LUCA \citep{chen_identification_1997}. A synthetic approach to ligate the minimal catalytic unit of RNaseP \citep{waugh_design_1989}, labelled M1, to a guide sequence (GS), produced an M1GS which performs targeted cleavage \citep{derksen_rnase_2015}. Crucially from a computational and synthetic viewpoint, the M1GS approach enables complete \emph{in vitro} reactions. By drawing on the now large library of known RNaseP sequences, artificial \emph{in vitro} selection may explore an enormous space of GS targeting \citep{zou_engineered_2004}. 

\subsection{Automata theory}

Automata theory is the study of mathematical models of computation. It is important to recognise that the definition of computation used here goes beyond the design of computing devices to the mathematical formalisation of an algorithm as an effective procedure for performing a calculation \citep{turing_computability_1937}. Full reviews of automata and computational theory are found in the canonical texts by Hopcroft and Ullman \citeyearpar{hopcroft_introduction_1979} and Sipser \citeyearpar{sipser_introduction_2006}.

For our purposes in this Ansatz, we briefly establish that each model of computation, or automaton, is defined as an \textbf{n-tuple}, meaning it is composed of \textbf{n} distinct components. To illustrate by example, an automaton called a finite state system is a 3-tuple $(Q, \Sigma, \delta)$ where $Q$ is a set of possible internal states of the system and $\Sigma$ (sigma) is the set of possible distinct inputs to the system. The internal state of the system may change according to a given transition function $\delta$ (delta), which maps the current state to a new state dependent on the observed input. This mapping is written as $Q \times \Sigma \rightarrow Q$. We observe the state, and consider this to be the output of the system in response to the input. In this conceptualisation, we imagine the system transitioning within a space of all possible configurations of the state, occupying one of these locations at any given time. The state transitions occur in discrete steps, meaning it is always at a single point, never in between points. Looking at our system during its journey through this space of all possible configurations, we may say that the system state is determined by the past inputs which in turn guides the next transition in response to input.
In this Ansatz, we will be applying standard constructions from automata theory, imagining how they might be instantiated with RNA polymers.  

\section{Definitions}

\paragraph{RNA enzymes} The steps of the computation, referred to as transitions of the automaton, consist of modular additions and subtractions to the RNA polymer(s) that represent components in the automaton. The reactions allowed are ligation, the joining of RNA polymers, and cleavage, the dissociation of an RNA polymer into parts. It is assumed that a desired ligation or cleavage RNA enzyme is available for any given target RNA polymer(s).

\paragraph{States, symbols and stacks} The state of the automaton is represented by a designated RNA polymer, termed the \textbf{state polymer}. The sequence of this polymer represents the current state of the automaton at any given time during the computation. The input to our RNA automata will be the sequential presentation of designated RNA polymers, termed \textbf{symbol polymers}. These  polymers are not enzymes, rather they represent symbols in an alphabet defined within the automata. The sequential presentation will be referred to as the \textbf{input word}. In the second and third iteration of our RNA automata, we will introduce an extensible memory in the form of a stack for storing symbol polymers. The symbols may be added and subtracted from the stacks by the actions of the ligation and cleavage enzymes in the same manner as for the state polymer. All of the modifications to the state and to the stack(s) are modular operations.

\paragraph{Assumptions} There are multiple possible implementation paths for the automata constructions explored here. Rather than an exploration of experimental design, our aim was to propose constructions that place RNA dynamics within the class of automata. The goal is for the reactions proposed to be the simplest possible reactions, exemplifying an idealised `perfect world' reaction process. In the real world such reliability is possible, but cannot be easily achieved due to environmental noise, off-specific reactions (where a small percentage of reactions occur on targets that resemble the desired substrate), concurrent non-sequential reactions, and reverse reactions amongst other sources of variation. 
It is therefore probable that a benchtop implementation of these automata, or an example of an automaton in early life, would require more nuanced designs. Importantly, while these more nuanced designs may involve more reaction steps or more components, these would be following the same idealised reaction process, embodying the computational dynamics.  

To that end we make the following assumptions:

\begin{itemize}
   
    \item The reaction volume is imagined to consist of RNA molecules suspended in an aqueous solution.
    
    \item For polymers representing the given alphabets, we assume that it is possible to generate the corresponding sequences with sufficient stability in order to fulfil the role of unique substrates (i.e., non-enzymatic polymers). 
    
    \item Each RNA polymer enzyme initiates only a single reaction.
    
    \item All possible reactions are assumed to go to completion (i.e., the reactants are used up completely).
    
    \item The reactions of the transitions do not generate reverse, off-specific reactions or reactions triggered by environmental noise.

    \item Input polymers are made available to the automaton in a modular and sequential manner, occurring as required in the sequence of transitions.

    \item A mechanism exists that makes input polymers available to the automaton in a sequential manner. This ensures that at the start of each new transition, a single polymer is drawn from the given sequence of input polymers, and made available to the automaton.

    \item For any given ligation and any given cleavage reactions, there exists an RNA polymer enzyme to initiate this reaction, not conditional on previous reactions. That is, for any RNA polymers $a$ and $b$, there exists an RNA polymer enzyme $x$ that ligates $a$ to $b$. Similarly, for any RNA polymer $c$, there exists an RNA polymer enzyme $y$ that cleaves $c$ into given sub-polymers $d$ and $e$. 
    
    \item  Between transitions, the reaction volume is in an inert state, prior to the introduction of a new input which marks the start of a new transition.
    
    \item Transitions do not require the resolution of `race conditions', where the order of possible reactions at the start of a transition may influence the configuration after the transition. For example, if an input polymer is both modified and placed on a stack, these reactions can occur in either order to produce the same outcome.
    
    \item Stack polymers are distinguishable to the automaton as modular units, i.e. there is a signal to indicate the beginning and end of stack polymers. When multiple stacks are utilised, the automaton can distinguish between the stacks.
    
    \item The reactions profile of the transitions can proceed without consuming an input polymer from the input stream or without cleaving a stack polymer, or without ligating a polymer to the stack.
    
    \item Accept and reject states are designated as specific polymers before automaton construction (further detail below) and are assumed to be distinguishable by an external observer.
    
\end{itemize}

\section{RNA automata}
An automaton is an abstract construction for performing a computation. We will start with a \textbf{finite automaton (FA)} in which only the state polymer is modified in response to the input word. We will then iterate to add one and two stack polymers. At each automaton type, we will first give a theoretical background and notation from automata theory. We then outline the construction of the given RNA automaton and give worked examples of a computation. 

\subsection{Finite Automata} 
\paragraph{Background} A finite automaton (FA) progresses through sequential transitions, where the state may change in response to the input. The transitions are carried out with reference to a defined set of transition rules for moving between any given state in response to the input. Certain states may be designated to have meaning with respect to the input word, e.g. an Accept or Reject state may be reached and, if halted on, signify a response to the total input. 

A FA is defined by a 5-tuple, $ (Q, \Sigma, \delta,  q_0,  F)$. $Q$ is the finite set of states that the automaton may visit. $\Sigma$ is the alphabet that the input may be drawn from. $\delta$ is the transition function, the rules for moving between states, of the form $Q \times \Sigma \rightarrow Q$. $q_0$ is a designated starting state, where $q_0 \in Q$. $F$ is the set of accept states, where $F \subseteq Q$.

\paragraph{RNA-FA components} The transition function holds the instructions for manipulating the state in response to the current input. Rules within the transition function take the form $(q_i, a) \xrightarrow{R} (q_j)$, which means that, for automaton R, if the current state is $q_i \in Q$, and the current input is a letter of the alphabet $a \in \Sigma$, then the automaton will change state to $q_j \in Q$. In an RNA-FA, the transition rules are embodied in the reaction profile of RNA enzymes. In the example, when the state polymer has sequence $q_i$, and the symbol polymer with sequence $a$ is the current input, a ligation or a cleavage reaction occurs to the state polymer such that it is lengthened or truncated to become the sequence $q_j$. 

The computation of the automaton proceeds in a series of steps with defined stages, starting from an inert point either at the initialisation of a new automaton or after the conclusion of a previous transition. First, a symbol polymer from the input is introduced, and may be recognised as a pair with the current state polymer sequence $q_i$. A stage of reactions occur to completion, which may alter the state polymer and thus change the state of the automaton. A final `cleanup' stage resets the reaction volume to an inert state, prior to the introduction of a new input which marks the start of a new step. 

\paragraph{RNA-FA notation}  RNA enzymes in our automata perform ligation or cleavage reactions, which are denoted by $\lambda$ (lambda, ligation) and $\mu$ (mu, cleavage) respectively. The first term in a recognition pair is the subject of the reaction that will be ligated to or cleaved from. For ligation, the second term in the recognition pair is directly ligated to the first or acts as a catalytic element to facilitate modification of the first term. For cleavage, the second term is a catalytic element. 

For ligation in the RNA-FA:
\begin{center}
\begin{equation}
\lambda(x, y): Q \times \Sigma \rightarrow Q
\end{equation}
\end{center} i.e. $z=\lambda(x, y)$ where $z$ is the state polymer such that $z$ is the ligation of $x$ with $y$ or catalysed by $y$. 

Similarly, for cleavage in the RNA-FA: 
\begin{center}
\begin{equation}
\mu(x, y) : Q \times \Sigma \rightarrow Q 
\end{equation}\end{center}
i.e. $z=\mu(x, y)$ where $z$ is the state polymer such that $z$ is the cleavage of $x$ catalysed by $y$.

\begin{SCfigure}[1]
    \centering
    \caption{Illustration of $\lambda, \mu \text{ and } \kappa$ transitions. Top: The $\lambda$ transition is a ligation reaction which creates a 3’5’ phosphodiester linkage between the state polymer and the input polymer to form a single polymer. The sequence of the new polymer corresponds to the state $Q_1$. Middle: The $\mu$ transition is a cleavage reaction which separates the state polymer. The sequence of the truncated state polymer corresponds to the state $Q_0$. The remaining polymer from cleavage is degraded or washed out prior to the next transition. Bottom:  The $\kappa$ transition in which no enzymatic reaction takes place. The state polymer and the input polymer are not recognised as a template and do not catalyse a reaction. The input polymer is degraded or washed out prior to the next transition.}
    \includegraphics{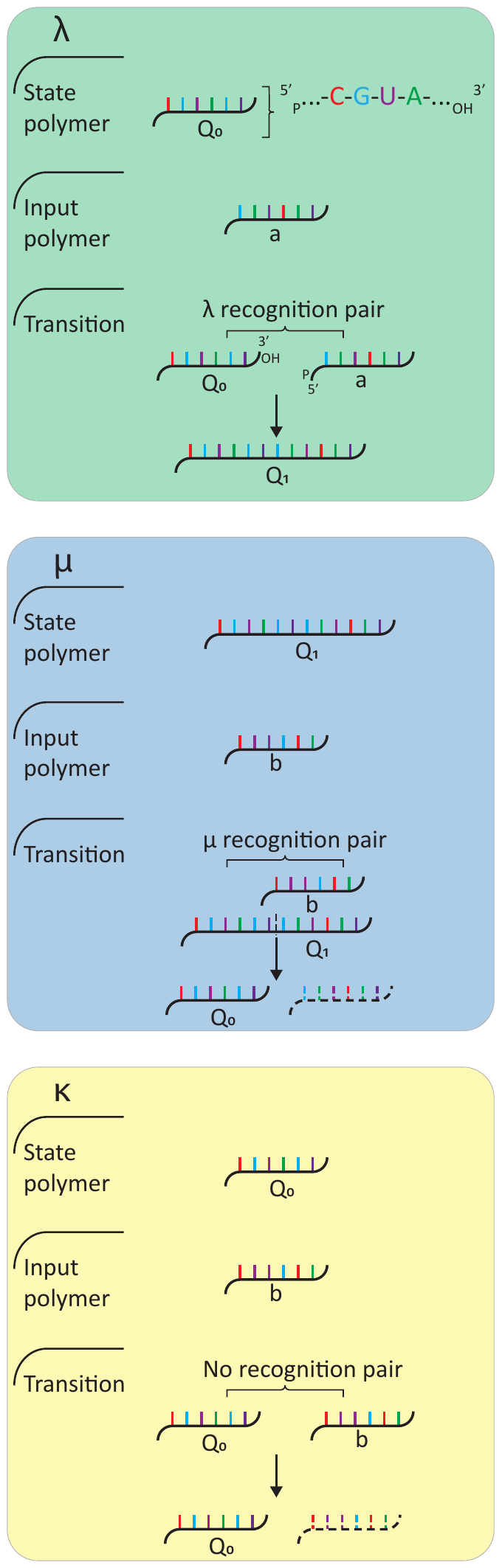}
    \label{fig:transitions}
\end{SCfigure}

We also define a stasis operation, where the response to an input is to remain in the same state:  
\begin{center}
\begin{equation}
\kappa(x, y) : Q \times \Sigma \rightarrow Q
\end{equation}\end{center}
i.e. $x = \kappa(x, y)$ for all $y$. The transitions are illustrated in Figure \ref{fig:transitions} as an accompaniment to the worked example of the RNA-FA.

\subsubsection{RNA-FA for \texorpdfstring{$b^*(ab^+)^*$}{b*(ab+)*}} 
To perform a computation, we will encode the required alphabet for our automata into unique symbol polymers drawn from the ribonucleotide ACGU alphabet with minimum length determined as required for RNA enzyme activity. We will also encode the unique initial state polymer in the same format. We then design RNA enzymes that will transition the state polymer through the designated state sequences in the presence of the symbol polymers. 

To illustrate, the RNA-FA we are constructing is to determine whether a specific ordering of symbol polymers, the input word, conforms to a pattern. Our RNA-FA will  recognise input sequences of the form $b^*(ab^+)^*$. The $^*$ indicates `0 or more of' and the $^+$ operator indicates `at least 1 of'. Put together, $b^*(ab^+)^*$ indicates the input polymer may have an arbitrary arrangement of $b$'s but any instance of $a$ must followed by at least 1 $b$. An empty sequence, or a sequence consisting only of $b$'s should be accepted by this definition. A pair of specific RNA polymers will represent $a$ and $b$, forming $\Sigma$, from which an ordering of such polymers is chosen as the input word. The reactions cascading from the sequential presentation of the input of the form $b^*(ab^+)^*$ will result in reaching (or remaining in) a sequence of the state polymer designated as the accept state, and any non-conforming input words will reach a reject state. At the exhaustion of input, the sequence of the state polymer determines the acceptance or rejection of the input sequence.

The RNA-FA is represented as a five-tuple, $(Q, \Sigma, \delta, q_0,  F)$, where: \newline
$Q \text{ is the set of states } \{q_0, q_1, q_2\}$ where each $q_i$ is a unique sequence of the state polymer. \newline
$\Sigma \text{ is the alphabet } \{a, b\}$ where $a$ and $b$ are unique symbolic RNA polymers. \newline
$F \text{ is the set of accept states } \{q_0\}$ \newline
The transition function $\delta$ is given by the following transitions:
\[ \delta =\begin {cases} 
						(q_0, a) = q_1 \equiv \lambda(q_0, a) 				\\
                        (q_0, b) = q_0 \equiv \kappa(q_0, b) 				 \\
                        (q_1, a) = q_2 \equiv \lambda(q_1, a)  		          \\
     					(q_1, b) = q_0 \equiv \mu(q_1, b) 					   \\ 
     					\end{cases}
 \]
\begin{figure}[H]
\centering
    \includegraphics[]{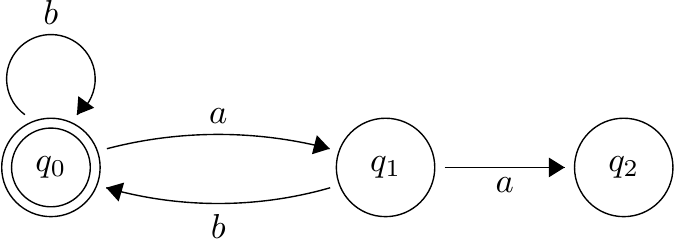}
    \caption{State diagram for the RNA-FA.}
    \label {fig:RNAFA}
\end{figure}

\newpage
If we take as input \emph{bab}: 
\begin{enumerate}
\item With state polymer sequence $q_0$, $b$ is a stasis symbol polymer.
\item With state polymer sequence $q_0$, $a$ is recognised and the state polymer is ligated to form the sequence $q_1$.
\item With state polymer sequence $q_1$, $b$ catalyses cleavage of the state polymer to return to the sequence $q_0$.
\item At the exhaustion of input, the state polymer has sequence $q_0$, so the automaton \textbf{accepts}.
\end{enumerate}

If we take the input \emph{aba}: 
\begin{enumerate}
\item With state polymer sequence $q_0$, $a$ is recognised and the state polymer is ligated to form the sequence $q_1$.
\item With state polymer sequence $q_1$, $b$ catalyses cleavage of the state polymer to return to the sequence $q_0$.
\item With state polymer sequence $q_0$, $a$ is recognised and the state polymer is ligated to form the sequence $q_1$.
\item At the exhaustion of input, the state polymer has sequence $q_1$, so the automaton \textbf{rejects}. 
\end{enumerate}

\paragraph{RNA-FA computations} 
We may ask what kind of computing tasks could such an RNA-FA perform? We may observe that at any given point during the computation, the current sequence of the state polymer describes a trajectory of visited states and inputs encountered. If the input word conforms to an accepted pattern, the RNA-FA will step through an accepting path. Formally, a FA may process the class of \textbf{regular languages}. Regular expressions, which describe regular languages, can  specify \emph{patterns} used in searching operations~\citep{sipser_introduction_2006}. At the molecular scale, biology makes prolific use of regular expressions. In particular, the non-coding subset of the genome contains an enormous variety of patterns, referred to as motifs, which characterise families of genomic elements. For example, the upstream promoter region of a gene can be described as a regular expression such as $G[x]AT[x]AA[x]AT[x]CA$, where $[x]$ represents any of the nucleotides [AGCT] (in this case for the bacterial gene argR  \citep{mcguire_conservation_2000}). Identifying this phenomenon led to significant progress in the practice of scanning and annotating the genome for motifs \citep{brazma_predicting_1998,brazma_approaches_1998, das_survey_2007}. The capability to perform FA computations confers a powerful pattern recognition ability to this simple arrangement of RNA polymers. 

A limitation of a FA is that any instance of returning to a previously visited state effectively erases the encoding of the trajectory beyond that state and as such the FA cannot maintain an extensible memory of repeated input. In other words, if a loop exists or the FA may return to some earlier state, there is no way to encode the number of times a loop has been traversed or a given state visited. In the next automaton, we will augment our RNA-automata with a polymer to serve as extensible memory.

\subsection{Push-down Automata}

\paragraph{Background}
To implement a memory component in our RNA automata, we will introduce RNA polymers with a purely symbolic, informational role. The automaton structure will be a push-down automaton (PDA) which operates along the same principles as the FA with states and transition rules. A PDA is augmented by the addition of a stack which can encode information over the course of the computation in an extensible manner. The stack operates according to a `Last In First Out' principle, in which symbols are prefixed to the top of the stack in a `push', and removed from the top of the stack in a `pop'. 

A PDA is defined by a 7-tuple, $(Q, \Sigma, \Gamma, \delta, q_0, Z_0, F)$. $Q, \Sigma, q_0$ and $F$ are defined as above for the FA. With the addition of a stack, we now include $\Gamma$ (gamma) as a finite set constituting the stack alphabet, and an initialising stack symbol $Z_0 \in \Gamma$. The PDA is defined with the empty symbol $\epsilon$ (epsilon). The input is defined as $\Sigma_\epsilon \equiv \Sigma \cup \{\epsilon\}$, in which $\epsilon$ may appear in place of input. When $\epsilon$ appears in the input, the transition may occur without reading a symbol from the input and without progressing to the next input symbol. $\Sigma_\epsilon$ is required for the full power of a deterministic PDA \citep{autebert_context-free_1997}. The stack input and output are defined as $\Gamma_\epsilon \equiv \Gamma \cup \{\epsilon\}$, in which $\epsilon$ may appear in place of the top stack symbol. When $\epsilon$ appears in the stack input place, the transition may occur without reference to the symbol on the top of the stack. When $\epsilon$ appears in the stack output, the transition proceeds without a symbol being placed on top of the stack.  

The transition function $\delta$ is of the form $Q \times \Sigma_\epsilon \times \Gamma_\epsilon \rightarrow Q \times \Gamma_\epsilon$,  for example, $(x, y, u) \rightarrow (z, w)$, where  $x, z \in Q$, $y \in \Sigma_\epsilon$, and $u, w \in \Gamma_\epsilon$.
For clarity, we will make use of the instantaneous description notation \citep{hopcroft_introduction_1979}. In this notation the automaton has a configuration (i.e., instantaneous description) which is the tuple of the current state, remaining input and current stack contents. A transition from configuration $\delta (x, yL, uS)$ to configuration $(z, L, wS)$ is indicated by symbol $\vdash$ so that $(x, yL, uS) \vdash (z, L, wS)$. This means that during the transition from state $x$ to state $z$, the first input symbol $y$ of the input $yL$ is `consumed', and the top symbol $u$ of the stack $uS$ is replaced by symbol $w$, forming the new stack $wS$. Here $L \in \Sigma^*$ and $S \in \Gamma^*$, with $\Sigma^*$ and $\Gamma^*$ being the Kleene star of the input alphabet and stack alphabet respectively, which is the smallest superset containing all possible words derived from symbols in the input or stack alphabets, including the empty word. Incorporating $\epsilon$, in the step-relation $(x, yL, uS) \vdash (z, L, wS)$, $y, u$ and $w$ may be $\epsilon$. To maintain our automaton in a deterministic mode, we establish the rule that if a configuration $\delta(x, y, u)$ containing $(z, w)$ exists, then the configurations $\delta(x, \epsilon, u)$ and $\delta(x, y, \epsilon)$ are empty. Similarly, a configuration $\delta (x, y, u)$ may contain only one of $(z, w)$ or $(z, \epsilon)$.

\paragraph{RNA-PDA components} To realise a PDA we need to initialise an RNA polymer to operate as our stack. In overview, such a polymer would be a modular structure, consisting of symbol polymers drawn from the available alphabet $\Gamma$. Prefixing of a new polymer to the stack and popping the top polymer from the stack is carried out by the same class of ligating ($\lambda$) and cleaving ($\mu$) enzymes as are already in use. Interpreting the $\epsilon$ in terms of the RNA-PDA, this means that the reaction profile of the transition does not include the respective $\epsilon$ component. For input, this means that the reaction profile is such that the reaction does not consume the input polymer from the input stream. For the stacks, this means that the transition reaction ignores the polymer on the top of the stack, and for the stack output this means that the transition reaction will not result in a polymer being ligated to the stack. In other words, distinct reaction profiles can differentiate between these separate kinds of transitions, instead of employing an explicitly designated polymer $\epsilon$ to ensure that input stream, or top of the stack, are not consumed, or that stack contents are not updated. Shifting the design `burden' between reaction profiles and specialised symbols is characteristic of the program-data duality.

Finally, we will introduce special symbol polymers to indicate the end of the stack or of the input word. The stack is initialised with a special end-of-stack symbol polymer denoted by $\eta$ (eta). All input words, including the empty word, have a special end-of-input symbol polymer in the last position, denoted by $\nu$ (nu).

\paragraph{RNA-PDA notation} 

PDA operation proceeds by two independent modifications of the automaton state and the stack. For simplicity, we use a general notation $\circ$ to represent any of the reaction functions, i.e., ligation, cleavage and stasis, that is,  $\circ \in \{ \lambda, \mu, \kappa \}$. Consider a transition $\delta$ given by the instantaneous description $(x, yL, uS) \vdash (z, L, wS)$. The state modification produces new state $z = \circ (x, \circ (y, u))$, given a suitable reaction between polymers $y$ and $u$ resulting in $\circ (y, u)$, and followed by another suitable reaction between polymers $x$ and the intermediate result $\circ (y, u)$. This sequence is ensured by the Ansatz assumptions (e.g., each RNA polymer enzyme initiates only a single reaction, all possible reactions are assumed to go to completion, reactions do not generate reverse and off-specific reactions, and are not triggered by environmental noise, etc.). The stack modification produces new stack $wS = \circ(u, \circ(y, x))$, given a suitable reaction between polymers $y$ and $x$ resulting in $\circ (y, x)$, and followed by another suitable reaction between polymers $u$ and the intermediate result $\circ (y, x)$. Again, this sequence is ensured by the Ansatz assumptions. These two modifications eliminate the need for a reaction among three reactants: instead we need only two sequential reactions with two reactants each, using suitable enzymes. Importantly, the choice of reaction type $\circ (a,b)$ between two polymers $a$ and $b$ is determined when the automaton employing this reaction is designed, that is, we do not expect this type to be discovered during the transition as a result of some additional conditions.

\paragraph{RNA-PDA computations} 

To construct an RNA-PDA, we have added a memory component and augmented our transition rules to read and write from this memory. A stack allows the RNA-PDA to perform cardinality and one-one correspondence tests between distinct components of an input. More generally the stack allows for an extensible counter. The extensibility of the memory is crucial to the power of the RNA-PDA. An RNA-FA could be designed to determine any given \emph{finite} input word incorporating loops, however, each loop would require a distinct subset of states and therefore come at the cost of a significant expansion in the state complexity. As such, we may observe that expanding from an RNA-FA containing exclusively functional components to include a component which serves a purely informational role, in the form of a stack, represents a significantly simpler path to recognition of more complex patterns containing loops.  

In particular, computing with a stack has implications for encoding and decoding with an RNA-PDA. The process of writing and reading with the RNA-PDA can be formally captured within the \textbf{deterministic context-free languages} (DCFLs) \citep{sipser_introduction_2006}. The transition rules of an RNA-PDA can embody a deterministic context-free grammar (CFG) \citep{sipser_introduction_2006}. In general, the derivation of a string using a CFG requires intermediate strings which are stored on the PDA stack. Thus, the stack of an RNA-PDA can provide storage for intermediate polymers derived  from the input polymer. As such, an RNA-PDA operating as a parser may encode information from the environment into a persistent RNA form, or decode some previously encoded RNA information. A deterministic PDA based parser has been described (LR, \citep{knuth_translation_1965}) which could be implemented by the RNA-PDA. Additionally, the stack may also embody the computational result of an encoded or parsed string when an encoding/decoding PDA accepts an input. 

Encoding and decoding in RNA automata is the first time we can observe a distinction and a relationship between the functional nucleic acid components carrying out the automata operations, and the informational nucleic acid components. Through encoding, a history of states and inputs can be recorded and retrieved through decoding. There is an important limitation in this relationship. In the RNA-PDA this informational representation lacks a reflexive relationship in which the information itself becomes the subject of manipulation. Such a reflexive relationship emerges when there is a cross-reference, an ability to copy and compare within the encoding. Put simply, the automata requires an additional space for copying and manipulating a stored encoding in successive transitions. In the next iteration, we will further augment our RNA-PDA with an additional stack to permit this exploration.

\subsection{Turing machines and Two-stack pushdown automata}

\paragraph{Background}
A Turing machine (TM) is an abstract general-purpose computing device, introduced as a formal model of computation, and intended to capture the entire class of computable functions (i.e., `algorithms') \citep{turing_computable_1937}. As a computing device, the TM surpasses the capability of PDAs, being the most powerful computing model. 
A TM uses a finite set of rules (program) which modify symbols on an infinite tape (data), with the latter distinguished from the stack by being accessible at any location along the tape. The tape is split into discrete cells each capable of holding a single symbol, or being blank. The TM is conceptualised to have a `read-head' positioned at one cell of the tape, at which it may read and modify the symbol. At then end of each transition, the read-head may move one cell to the left or right along the tape. Formally, a TM is defined as a 7-tuple, $(Q, \Sigma, \Gamma, \delta, q_0, q_{acc}, q_{rej})$ where Q, $\Sigma$ and $\Gamma$ are defined as the set of states, the input alphabet and the tape alphabet respectively. $q_0$ is the starting state and $q_{acc}$ and $q_{rej}$ are the predetermined accept and reject states. The $\delta$ transition function is defined as $\delta : Q \times \Gamma \rightarrow Q \times \Gamma \times \{L, R \}$ where $L$ and $R$ represent left or right movement of the read head. Mechanically, the tape and read-head operation of the TM is a departure from the construction of the RNA automata in this study. For this reason, we will implement a TM-equivalent automaton, the deterministic two-stack PDA (2PDA).  As the name suggests, a 2PDA may operates as the PDA above, with the addition of a second stack. Both stacks may be accessed during a transition, including switching a symbol between the stacks. The equivalence of a 2PDA to a TM may be demonstrated by simulation. A 2PDA may simulate a TM by assigning the two stacks to represent the portion of the tape to the left and right of the read-head. Interestingly, information storage on the two-stacks of a PDA is orthogonal to information storage via states, and it can be demonstrated that every TM has an equivalent deterministic, single-state 2PDA \citep{koslowski_deterministic_2013}.\newline Standard constructions of PDAs and 2PDAs include a $\Sigma$ input source external to the automaton. By contrast, TM construction incorporates the input as a buffered tape, where $\Sigma \in \Gamma$. These are not fundamental differences, and it can be observed that both constructions can be made equivalent by explicitly designing a TM to buffer an input word from an external source prior to computation, or to buffer the stack of a 2PDA with the input word \citep{koslowski_deterministic_2013}. In the RNA-2PDA, we will buffer the stack with the input word prior to computation.

A 2PDA is defined by a 7-tuple, $(Q, \Sigma, \Gamma, \delta, q_0, Z, F)$. We will refer to the two stacks as the `left' (L) and `right' (R) stack and we will initialise both stacks with $Z$, where $Z = \Gamma_L^* \times \Gamma_R^*$. As we will be initialising our 2PDA with the input on the L stack, the input alphabet $\Sigma \in \Gamma$.  The $\delta$ function is of the form $\delta: Q \times \Gamma^L_\epsilon \times \Gamma^R_\epsilon \rightarrow Q \times \Gamma^L_\epsilon \times \Gamma^R_\epsilon  $, with $\epsilon$-transitions as above.

\paragraph{RNA-2PDA components} To realise a 2PDA we will need to add a second stack polymer, initialised with a unique end-of-stack symbol. To enable the selective popping from and pushing to each stack, we will assume that, within the automaton, a symbol polymer on the left stack will be distinguishable from a symbol polymer on the right stack. 

\paragraph{RNA-2PDA notation} As we now have two stacks, we will re-define the end-of-stack symbols to be end-of-left, $\nu$, and end-of-right, $\eta$.

Possessing two stacks which can serve as storage and processing space enables an automaton to perform repeated computation on iterations of intermediate results. A worked example of an RNA-2PDA is given in appendix B for $a^{2^n}$.

\paragraph{RNA-2PDA computations} 

The languages computable by a TM or 2PDA are referred to as the \textbf{recursively enumerable languages} \citep{sipser_introduction_2006}. The recursion theorem demonstrates that an automaton which can read non-destructively and perform copying and comparison operations, such as a 2PDA, can derive its own description and compute with it \citep{sipser_introduction_2006}. Specifically, if an RNA-2PDA, R, was designed to utilise recursion, then it may encode its own transition rules [R] onto the stacks. The aim may be to self-reproduce, or R may go on to perform any 2PDA computation on [R] and any other input.

To consider this dynamic in our class of RNA-2PDA automata, recall that we considered the single stack of the RNA-PDA to represent a non-reflexive relationship between the functional components and the informational components encoding the history of states and inputs encountered. Such stored information could modify the progression of a subsequent computation but not itself become the subject of manipulation. In the RNA-2PDA, the second space for copying and comparison allows this information to be manipulated. The recursive ability of the two-stack automaton to encode its own transition rules on the stacks means that these symbolic RNA polymers may be copied and modified as with any other polymers on the stacks. The automaton may be represented in both a functional and informational form, with encoding and decoding allowing a reflexive relationship between the representations.

Importantly, the RNA-2PDA may hold an informational representation of any RNA enzyme automaton in the form of an abstracted representation of the internal relationships of the automaton, i.e. as an alphabet and transition rules. Such an encoding contains the information necessary for testing inputs and modifications of the encoding at the level of the encoding. In other words, some suitable automaton may \emph{simulate} the computation of itself or another automaton from such an encoding. In fact, the ability to match and substitute across two stacks enables the RNA-2PDA to simulate \emph{any} automaton through universal computation.

\subsection{Universal computation with an RNA-UPDA}

\paragraph{Background}
A universal automaton, such as the canonical Universal Turing Machine (UTM) \citep{turing_computable_1937}, may read the description of any automaton and simulate that automaton on some input. There are three key components to universal computation; the first is generating the description of the automaton to be simulated by conversion of its transition rules and inputs into the alphabet of the universal automaton. The resulting description is referred to as an \textbf{encoding}. Second, the simulating automaton, in this case our RNA-UPDA, must be capable of manipulating the encoding to faithfully simulate computation of the encoded automaton. Third, this simulation must be facilitated within a memory layout (on tape(s), stacks etc.) that accommodates the encoding and the inputs and outputs of the simulated automaton. We will demonstrate an RNA-UPDA encoding strategy and procedures for handling the encoding, and explore how a simulated computation may be accommodated in a specialised RNA-UPDA 3-stack construction.

To show capability of the RNA-UPDA for universal computation, we will focus our attention on the known set of small, size efficient UTMs  \citep{shannon_universal_1956, woods_complexity_2009}. One strategy for the implementation of a size efficient UTM is to simulate 2-tag automata. 2-tag automata are a member of the $m$-tag automata \citep{minsky_size_1962}, which compute by modification of a word on a single linear tape. An $m$-tag automaton always reads the first symbol of the input word, deletes $m$ symbols from the start of the input word and then appends some symbol(s) to the end of the word. Importantly, any algorithm that can be computed by a TM-equivalent automaton can be computed by a 2-tag automaton \citep{cocke_universality_1964}. Formally, the 2-tag automaton is given by the 2-tuple $(S, T)$ where $S$ is the alphabet of unique symbols that may be read from or appended to the word being computed for $S = \{s_1, s_2, \ldots, s_n, s_{n+1}\}$ in which $s_{n+1}$ is a halt symbol. The transition rules $T$ map the members of $S$ to the finite set of words $S^*$, which are appended to the input word. A transition rule is of the form: $s_i \rightarrow \alpha_i$ for $i=\{1 \ldots n\}$ where $\alpha_i = s_{i_1} s_{i_2} \ldots s_{i_l}$ and in every transition two symbols are to be removed from the start of the word.

We will aim to demonstrate that our RNA-UPDA can simulate any 2-tag automaton on any input. First we will demonstrate an example UPDA alphabet into which the transition rules and input of any target 2-tag automaton may be encoded to be simulated by the RNA-UDPA. Second, we will describe RNA-UPDA functions for matching and copy operations that are required to carry out the simulation. Finally, we will describe an RNA-UPDA with 3 stacks that may simulate any target 2-tag automaton by manipulating the encoded alphabet. We aim to demonstrate that an RNA-UPDA may simulate any 2-tag automaton by demonstrating that the simulated 2-tag input word during computation and after halting are in concordance with that which would be observed in the target 2-tag automaton.

\paragraph{RNA-UPDA components} 

To first encode the finite alphabet $S$ of the target 2-tag automaton into the alphabet of the RNA-UPDA, we will encode each symbol of $S$ as complementary pairs of nucleic acid polymers.

\paragraph{Alphabet to symbol polymer encoding}
Every $s_i \in S$ is assigned a pair of complementary symbol polymers denoted $a_i$ and $\bar{a}_i$. Complementary refers to the property that for all $j$, the nucleotide at each position $a_{i_{j}}$ is matched at $\bar{a}_{i_{j}}$ by the complementary nucleotide to which it preferentially binds (i.e. $C \leftrightarrow G$ and $U \leftrightarrow A$). 

The alphabet of the target 2-tag automaton ($S$) is therefore encoded into the RNA-UPDA alphabet $A = \{\{a_1, \bar{a}_1\}, \{a_2, \bar{a}_2\}, \ldots, \{a_n, \bar{a}_n\}, \{a_{n+1}, \bar{a}_{n+1}\}\}$, where the symbol polymer $a_{n+1}$ represents a halt symbol.

\paragraph{Input word to input polymer encoding}
The input to the RNA-UPDA consists of a word, denoted as $K$, composed of the $a_i$ members of the encoded pairs of symbol polymers in the RNA-UPDA alphabet $A$. Put together, the input of the RNA-UPDA will be the modular polymer $a_{K_0} a_{K_1} a_{K_2} \ldots a_{K_n}$.   

\paragraph{Transition rule to instruction polymer encoding}
The polymers which will serve as instructions for modifying the input, denoted as $D$, will be composed of the $\bar{a_i}$ members of the encoded pairs of symbol polymers in the RNA-UPDA alphabet $A$. The special symbol polymer $\triangleright$ will serve to demarcate the start and end of each instruction polymer. It will be a requirement of encoding that every pair $\{a_i, \bar{a}_i\} \in A$ is associated with exactly one instruction polymer $d_i \in D$ \citep{rogozhin_small_1996}. Put together, every $d_i \in D$ for $i = 1 \ldots n$ is of the form $\triangleright \bar{a_i} \bar{\rho}_i$ where $\bar{\rho}_i = \bar{a}_{i_0} \bar{a}_{i_1} \ldots \bar{a}_{i_l}$ for $l \geq 1$.

To initialise the stacks for our RNA-U2PDA, we generate instruction polymers of $D$ and push each to the `instruction' stack, with end-of-stack symbol $\nu$. A special symbol polymer $\bigtriangleup$ will be initialised at the top of the `instruction' stack and will serve to demarcate the boundary of the instruction polymers and the input polymers. For input, we generate the input polymers of $K$ and push each to the `input' stack, with end-of-stack symbol $\eta$. We also initialise a third `working' stack, with end-of-stack symbol $\omega$. This stack will take part as a temporary holding space during computation.

\begin{center}
    \begin{displaymath}
        \text{Instruction}\left\{  
        \begin{array}{l} 
        \bigtriangleup \\  
        \left.
            \begin{array}{l} 	d_0 \\	d_1 \\	\vdots \\	d_n	\end{array}
        \right\} D \\
        \nu
        \end{array} \right.
        \left.
        \begin{array}{r}
	        A^*\left\{
            \begin{array}{r}	a_{K_0}\\	a_{K_1}\\ \vdots \\ a_{K_m}
    	    \end{array}
            \right.\\
        \eta
        \end{array}
        \right\} \text{Input}
    \end{displaymath}
    \begin{displaymath}    
        \text{Working}\left\{
        \begin{array}{l} \omega	
	    \end{array}
        \right\}
    \end{displaymath}
\end{center}

To simulate a 2-tag automaton transition, the RNA-UPDA first engages in \textbf{matching} (search, comparison) between the topmost input symbol polymer of the `input' stack at the start of the transition and the LHS of the instruction polymers, $\bar{a}_i$ on the `instruction' stack. After identifying the associated instruction polymer, the RNA-UPDA then engages a \textbf{copying} procedure to generate $\rho_i$ from $\bar{\rho}_i$ on the RHS of the instruction polymer. To enable these functions during the computation, we will utilise the binding between complimentary polymers as a targeting mechanism, which is further outlined below.

\paragraph{Matching}
 The matching function in the RNA-UPDA simulates the transition rule lookup of the 2-tag automaton. When the matching function is invoked, it takes as input a single symbol polymer located on the top of the `working' stack (denoted $a_{match}$), identifies an associated instruction polymer on the `instruction' stack and initiates the copy function. The matching function consists of a repeating cycle:
 \begin{itemize}
     \item $a_{match}$ is popped from the `working' stack and allowed to bind to the LHS of the topmost instruction polymer of the `instruction' stack.
     \begin{itemize}
         \item If the current instruction polymer $d_i$ is associated with $a_{match}$, then the complementary binding $a_{match} \leftrightarrow \bar{a_i}$ will serve as the initiating signal for the copy function of the RNA-UPDA and $a_{match}$ is discarded.
         \item If no binding occurs, $a_{match}$ is pushed back to the `working' stack and the topmost instruction polymer is temporarily cycled to the `input' stack.
     \end{itemize}
 \end{itemize}

\paragraph{Copying}
The copy function is the first part of a two-step process within the RNA-UPDA that simulates the 2-tag automaton step of appending a new symbol or word to the input word. The copy function takes as input an instruction polymer and outputs a new polymer to the `working' stack: 
\begin{itemize}
    \item The input to the copy function is $\bar{\rho_i}$, located on the RHS of $d_i$. $\bar{\rho}_i$ serves as a template for the process of \textbf{template-directed ligation} \citep{doudna_rna-catalysed_1989} in which short, random sequence polymers present in the reaction volume, but not encompassed within $A$, align to $\bar{\rho}_i$ through complementary binding.
    \item As short polymers align to $\bar{\rho}_i$, these are ligated together by RNA ligases into the complete $\rho_i$ polymer.
    \item The polymer $\rho_i$ is pushed to the `working' stack.
\end{itemize}

\paragraph{RNA-UPDA notation}

We will now demonstrate an example RNA-UPDA construction to simulate some 2-tag automaton. We will use a binary encoding, in which $C/G$ corresponds to 1 and $A/U$ corresponds to 0. This encoding will serve to illustrate the relationship of nucleotide encoding to binary representation, and to better illustrate the complementarity mechanism\footnote{For an instantiation, we assume that it is possible to generate sequences with no enzymatic activity within the RNA-UPDA and a maximum threshold for similarity that minimises off-target binding. For example, an encoding of each $s_i \in S$ to a random sequence (ACGU)* with lengths >18 nucleotides and <70\% identity between any two sequences would be in keeping with accepted oligomer design to maximise specificity. Within these bounds there remains a very large space of unique sequences.}. 

Given a 2-tag automaton with alphabet $\{ s_1, s_2, s_3, \ldots, s_n, s_{n+1} \}$ encode every $s_i \in S$ as:

\begin{center}
\begin{math}
(a_i, \bar{a}_i) \in A:
\begin{cases}
    \begin{aligned}
        &a_1 = C    &&1     &&\bar{a}_1 = G \\
        &a_2 = CA   &&10    &&\bar{a}_2 = GU \\
        &a_3 = ACC  &&011   &&\bar{a}_3 = UGG \\
        &a_4 = CAA  &&100   &&\bar{a}_4 = GUU \\
        &a_5 = CAC  &&101   &&\bar{a}_5 = GUG \\
        &a_6 = CCA  &&110   &&\bar{a}_6 = GGU \\
        &a_7 = ACCC &&0111  &&\bar{a}_7 = UGGG \\
        &a_8 = CAAA &&1000  &&\bar{a}_8 = GUUU \\
        &\vdots     &&\vdots    &&\vdots \\
        &a_{n+1} = \emph{halt} &&
        &&\bar{a}_{n+1}= \bar{\emph{halt}}
    \end{aligned}
\end{cases}
\end{math} \newline
\end{center}
The transition rules of the simulated 2-tag automaton $T$ will be encoded into the instruction polymers of the RNA-UPDA $D$. For example, let the 2-tag automaton transition rule $t_1$ be $s_1 \rightarrow s_6 s_7$. The encoded instruction polymer $d_1$ will be:

\begin{center}
    \begin{equation*}
        \underbrace{\triangleright}_{marker}
        \overbrace{G}^{\bar{a}_1}
        \underbrace{\sqcup}_{spacer}
        \overbrace{
            \overbrace{GGU}^{\bar{a}_6}
            \underbrace{\sqcup}_{spacer}
            \overbrace{UGGG}^{\bar{a}_7}}^{\bar{\rho}_i} 
        \underbrace{\triangleright}_{marker}
    \end{equation*}
\end{center}

The input word of the simulated 2-tag automaton will be encoded into the symbol polymers of the RNA-UPDA that make up the input polymer $K$. For example, let the 2-tag automaton input word be $s_1 s_3 s_5$. The encoded input polymer $K$ will be:
\begin{center}
    \begin{equation*}
        \overbrace{C}^{{a_1}} 
        \underbrace{\sqcup}_{spacer} 
        \overbrace{ACC}^{{a_3}}
        \underbrace{\sqcup}_{spacer}
        \overbrace{CAC}^{{a_5}}
    \end{equation*}
\end{center}

We will now demonstrate a transition of the simulated 2-tag automata which updates the input word. The first symbol polymer of $K$ will be popped and utilised in the matching function, the second symbol polymer of $K$ will be discarded. 

\begin{center}
    \begin{equation*}
        \underbrace{\overbrace{C}^{{a_1}}}_{\text{Input to matching}} 
        \bcancel{\overbrace{ACC}^{{a_3}}}
        \quad 
        \overbrace{CAC}^{{a_5}}
    \end{equation*}
\end{center}

In our example 2-tag automaton, the transition rule $t_1$ is the associated rule for the symbol $s_1$. The RHS of $t_1$ is $\alpha_1$, which is the symbol or word that will be appended to the input word to complete the transition. In the RNA-UPDA, the RHS of $d_1$ is $\bar{\rho}_1$ which is not the direct equivalent of $\alpha_1$. There is an extra step in the RNA-UPDA, in which $\bar{\rho}_1$ is the template for the template-directed ligation construction of $\rho_1$ during the copy function of the RNA-UPDA. $\rho_1$, the output of the copy function, is the equivalent of $\alpha_1$:  

\begin{center}
    \begin{equation*}
         \overbrace{
            \overbrace{CCA}^{{a_6}}
            \underbrace{\sqcup}_{spacer}
            \overbrace{ACCC}^{{a}_7}}^{\rho_i}
    \end{equation*}
\end{center}

To complete the transition, $\rho_i$ is appended to $K$, via the working stack.

\begin{center}
    \begin{equation*}
        \overbrace{CAC}^{{a_5}} 
        \underbrace{\sqcup}_{spacer} 
        \overbrace{CCA}^{{a_6}}
        \underbrace{\sqcup}_{spacer}
        \overbrace{ACCC}^{{a_7}}
    \end{equation*}
\end{center}

This completes the simulation of the 2-tag transition $s_1 \rightarrow s_6 s_7$.

\paragraph{RNA-UPDA procedure for simulating a 2-tag automaton transition}

For each simulated transition of the 2-tag automaton, the RNA-UPDA will progress through the sequence of transitions below to modify the encoded input word. At each new cycle, $a_{K_{2j}}$ will represent the symbol polymer at the top of the `input' stack for $j \geq 0$ where $j$ is the number of cycles completed. In overview, the RNA-UPDA computation proceeds as:

\begin{enumerate}

\item If the symbol $a_{K_{2j}} = \emph{halt}$ then the RNA-UPDA halts and the `input' stack constitutes the output of the RNA-UPDA. Otherwise:

\item $a_{K_{2j}}$ is popped from the `input' stack and placed on the `working' stack. The symbol polymer $a_{K_{2j+1}}$ is popped and discarded. This completes the 2-tag step of removing the first two symbols of the input word. 

\item The matching function is initiated with $a_{K_{2j}}$ as input at the top of the `working' stack. The matching function cycles through the instruction stack until the instruction polymer $d_i$ associated with $a_{K_{2j}}$ is at the top. The matching function initiates the copy function. This completes the 2-tag step of matching the first symbol of the input word to the associated transition rule.

\item The copy function is initiated with the associated $d_i$ at the top of the `instruction' stack. The copy function constructs $\rho_i$, which is pushed to the `working' stack. 

\item The `input' and `instruction' stacks cycle, sequentially popping the top symbol polymer of the `input' stack and pushing to the `instruction' stack until the end of the stack symbol polymer $\eta$ is reached.

\item When $\eta$ is the top symbol polymer of the `input' stack, $\rho_i$ is popped from the `working' stack and pushed to the `input' stack. Steps 4, 5 and 6 complete the 2-tag step of appending $\alpha_i$ to the input word.

\item The `input' and `instruction' stacks cycle in the reverse order, sequentially popping the topmost symbol polymer of the `instruction' stack and pushing back to the `input' stack. When $\bigtriangleup$, the symbol polymer which demarcates the boundary of $D$ and $K$, is at the top of the `instruction' stack the RNA-UPDA has completed one transition of the 2-tag automaton and reset for the next simulated transition.   
\end{enumerate}

The RNA-UPDA always starts a new simulated 2-tag transition by popping from the start of the encoded input word and appending $\rho_i$ to the last position of the `input' stack. If the `instruction' and `input' stacks are conceptualised as the left and right portions of a single linear structure, the region containing the encoded input word residing between $\bigtriangleup$ to the end-of-stack symbol $\eta$ is at all times concurrent with the input word of the 2-tag automaton being simulated. At the end of step 7, the `input' stack, excluding the end-of-stack symbol, is concurrent with the input word of the simulated 2-tag automaton. Hence, this simulates the target 2-tag automaton on target input, and therefore, shows universality.

It is instructive to consider which components of the RNA-UPDA above are serving the role of symbolic `data', and which components are performing an instructive role to guide the progress of the computation as `program'. From this perspective the role of nucleotide polymers may cycle between representing data as members of $K$ on the `input' stack and representing program as members of $D$ on the `instruction' stack. RNA represents a natural substrate for such \textbf{program-data duality} which we will examine further as a component of undecidability. 

\paragraph{RNA-UPDA computations}

In the above demonstration, universality was found by enacting a process of encoding the rules and input of automata into data for a simulating computation. An important consequence is observed if we recognise that the transition rules of the RNA-UPDA may be accessed by recursion which was introduced with the RNA-2PDA. These rules may be passed through the same encoding process as any other automaton. The encoded rules of the RNA-UPDA may then serve as the data input for a simulation of its own computation. Encoding an automaton into the data it computes generates an instance of \textbf{self-reference}. Such self-reference is a mechanism to generate the Liar paradox at the heart of undecidability, which we now turn our attention to.

\section{Undecidability}

The class of 2PDA automata are capable of generating undecidable statements, which can be exemplified by logical paradoxes like the Liar paradox, and leading to the halting problem \citep{turing_computable_1937}. To unpack the role of self-reference in generating undecidability in RNA automata, we will adapt the examples of Sipser \citeyearpar{sipser_introduction_2006} and the Liar paradox constructions of Prokopenko et al. \citeyearpar{prokopenko_self-referential_2019}. 

In the above UPDA, the automaton was constructed such that the output of the computation was the `input' stack after halting. For any 2PDA or equivalent, including the UPDA, there is a 2PDA that accepts or rejects the input. The input for an accepting automaton is formulated such that the question is answered by the accept or reject output. Suppose an accepting RNA-UPDA, $U$, into which we pass the encoding of some automaton $R$. With $R$ we will also pass $w$, the input word for which R accepts or rejects. $U$ will then accept or reject if $R$ would accept or reject. We may write $U = \{[R, w] \, | \, w \in \Sigma^* \}$. The [ ] notation indicates an encoding into a word of $\Sigma$ and we assume that special characters exist such that the encodings of the transition rules and the input word are distinguishable as such to $U$.

This relationship of $U$ to $R$ is:
\begin{center}
$U ([R, w]) \begin{cases}
    \text{ Accept if R accepts w}   \\
    \text{ Reject if R rejects w}   \\
    \text{ Run forever if R runs forever on w}
\end{cases}$
\end{center}
Observe that we can encode $R$ and pass this as the input word, $[\,R, \,[\,R\,]\,]$. In this case, R may accept or reject the word encoding itself or run forever. This relationship is not necessarily paradoxical, for that we will need a special automaton called a universal decider.

The proposed universal decider, $D$, is similar to $U$ above with the additional ability to reject when $R$ would run forever on $w$. The impossibility of the universal decider $D$ is demonstrated in the paradox that such an automaton generates. The relationship of $D$ to $R$ is:
\begin{center}
$D ([R, w]) \begin{cases}
    \text{ Accept if R accepts w} \\
    \text{ Reject if R rejects w or runs forever}
\end{cases}$
\end{center}

Now suppose a contrarian decider, $I$, which checks and inverts the relationship of $D$ to $R$ when the input word is $[R]$:
\begin{center}
$I ([R]) \begin{cases}
    \text{ Accept if R rejects [\,R\,] or runs forever} \\
    \text{ Reject if R accepts [\,R\,] }
\end{cases}$
\end{center}

The contrarian decider has introduced negation to the dynamics. If we now introduce self-reference to negation, we create an auto-negating paradox:
\begin{center}
$I ([I]) \begin{cases}
    \text{ Accept if I rejects [\,I\,] or runs forever} \\
    \text{ Reject if I accepts [\,I\,] }
\end{cases}$
\end{center}
We can write this as $I([I])$ will accept only when $I([I])$ rejects. 

This is a paradox pertinent to all computational frameworks capable of universal computation. In the preceding sections we have shown the theoretical construction of RNA automata that are capable of universal computation with self-referential dynamics. We may observe that RNA-automata demonstrate the key criteria of systems capable of demonstrating undecidable dynamics \citep{prokopenko_self-referential_2019}: RNA automata demonstrate \emph{program-data duality} as discussed above, \emph{access to an infinite medium} through a renewing supply of short RNA polymers and \emph{negation} through the ability to encode accept and reject representations which may be flipped.

\section{Discussion}

This Ansatz set out to probe the question of whether formal undecidability could be embodied in biological components. To do this we explored configurations of RNA polymers constructed into arrangements termed automata that compute functions on input. We aimed to formally express the RNA-mediated functions of ligation and cleavage in terms that aid in exploring automata construction. Within this framework, we surveyed a progression of RNA automata commencing with the purely functional construction of the RNA-FA, in which the automaton consists of only the RNA state polymer and the RNA enzyme polymers that carry out the transitions. An RNA-PDA was constructed by the addition of a Last-In-First-Out stack. The stack expands the automaton with a structure for storing RNA symbol polymers which may represent transient memory within the automaton. An RNA-2PDA was constructed by addition of a second stack. The RNA-2PDA is equivalent to a TM and can recognise the recursively enumerable languages. Automata in this class, including the RNA-2PDA, are able to reflexively encode a description of their program into data, and to compute with and instantiate this encoding. Here we first encounter clear program-data duality, turning a description of an automaton $M$ into some data $[M]$. From the foundation of a 2PDA, an RNA-UPDA was explored that could achieve universal computation, that is the capacity to simulate any other automaton in an encoded form. Universality enables such a system to explore the greatest possible solution landscape; but comes with the price of undecidable dynamics, e.g. when a universal automaton self-referentially runs on its own encoding. In other words, it becomes possible for such a system to generate computational undecidability, the outcomes of which may not be determined within the system itself. We have seen an example of such paradoxical `self-negating' computation, constructed in an analogy with the Liar paradox, which offers no possible resolution within its own set of rules. 

Our Ansatz is in two parts: 
\begin{itemize}
    \item[1] \emph{RNA automata can be constructed that embody computational models, up to Turing machine equivalence}.
    \item[2] \emph{At sufficient complexity (analogous to universal computation), RNA automata may generate self-reference and hence, computational undecidability. Continual resolution of computational undecidability represents a pathway to progressively expanded the boundaries and complexity of the automata, i.e. innovating.}
\end{itemize}

We have addressed the first part of the Ansatz above. To address the second part of the Ansatz, we pose a question that the prospect of undecidable biological computation raises. Since the paradox requires a perspective outside the system from which to observe and invert the output, where is the space in which the paradox may arise?

To answer the question we must ask if there exists a larger meta-system, encompassing the computational undecidability, which may play the role of universal decider and inverter. Such a system also contains the spark for removing the ceiling on biological complexity. This is because a key concept in undecidability as stated here is the lack of resolution for an automaton \emph{within its own set of rules}. Importantly, an undecidable problem is framed within a given formal system, and once the system is appropriately extended, the problem in point becomes decidable --- at the cost of generating other undecidable problems that inevitably arise in the extended system.

A well-known analogy of a meta-system which resolves computational undecidability \emph{at a given level} is an oracle machine, which supersedes a Turing machine \citep{turing_systems_1939}, being capable of deciding an outcome that could not be decidable by a universal Turing machine such as the RNA-UPDA. An oracle is some entity which is not itself a machine and which provides to a Turing machine some information from outside its own bounds. 
Turing gave a mechanical description of the interface of an oracle and TM of an o-machine to be a configuration of the o-machine in which the next state depends on feedback from the corresponding oracle \citep{turing_systems_1939}.

In context of sequential innovations, the o-machine concept was utilised by Penrose \citeyearpar{penrose_shadows_1994}, who defined the class of o-machines that may overcome the undecidable halting problem as the first-order o-machines. An o-machine comprises a TM and an oracle which is able to compute the values of a function which may not be computable. This combination, i.e., a TM and an oracle, has an expanded computational capacity relative to the TM alone. For example, a first-order $o$-machine comprises an oracle which can determine the value of the corresponding TM halting function. 
In turn there exist second-order $o$-machines with oracles capable of deciding halting states of the first-order $o$-machines, and so on, generalised in the concept of $\alpha$-order machines. Such a chain consecutively expands the boundaries of lower-order systems, by introducing a pointed \emph{innovation} (supplied by the corresponding oracle) in a form of a new description (i.e., axiom) added in  the  higher-order system. Importantly, the innovation can be provided to the system at the $(\alpha - 1)$-order as a form of feedback, reacting to which extends the system's boundary to construct an $\alpha$-order system (discussed further below).  

An o-machine, a Turing machine with an oracle, is analogous to augmenting the original logical system with a new, independent, axiom. From this basis, a continual, step by step process may follow in which the bounds of any individual logical system may be overcome and the system continually expanded, as suggested by Turing in his introduction on systems of ordinal logics (where a logic would now be described as a formal system) \citeyearpar{turing_systems_1939}: \begin{displayquote}``The well-known theorem of G\"{o}del (1931) shows that every system of logic is in a certain sense incomplete, but at the same time it indicates means whereby from a system L of logic a more complete system L$'$ may be obtained... A logic $L_{\omega}$ may then be constructed in which the provable theorems are the totality of theorems provable with the help of the logics $L, L_1, L_2, \ldots$''  \end{displayquote}

A continual, step by step process of expanding system boundaries in an attempt to `reconcile' a paradox, is a recurrent motif in studies of formal systems \citep{chaitin_algorithmic_1987, sayama_construction_2008, chaitin_life_2012, abrahao_paradox_2017}. An influential early result was established by Post from the perspective of the recursively enumerable sets, by stating that while no recursively generated logic is complete, \emph{every recursively generated logic may be extended} \citep{post_recursively_1944}. 
In doing so, Post showed that the complement set of the set of true propositions is not recursively enumerable, that is, the sets of propositions which can be `guaranteed' to be true, $T$, and false, $F$, do not exhaust the set of all propositions. The proposition which was shown to be outside of either of these two sets, i.e., an `undecidable' proposition, was constructed in a self-referential way, by recursively enumerating false propositions and identifying the set $S_0$ of corresponding positive integers. The incompleteness is shown by constructing the proposition describing the set $S_0$ itself: this proposition cannot be false (not in $F$) but has to remain outside of set $T$. It is precisely the addition of this proposition to the set $F$, making a new set $F'$, that constitutes the expansion of the logic (i.e., innovation), and so a sequence of such expansions/innovations may be developed. 

\subsection{An $\alpha$-order o-machine in biological automata}

In search for a meta-system to inform the biological automaton, the niche of the biological system is an obvious candidate. The inter-dependence of the niche and the genome is captured in the concept of the `reactive genome' \citep{gilbert_reactive_2003} here characterised by Griffiths and Stotz \citeyearpar{griffiths_genetics_2013}: \begin{displayquote}``The regulatory architecture of the genome reaches outside the genome itself, outside the cell, and outside the organism... Many of the factors involved in genome regulation are highly context-sensitive, which allows them to relay environmental information to a reactive genome which has evolved to let environmental input play an instructive role on the determination of phenotypes.''\end{displayquote}

If we recognise that the universal biological automaton is operating in an environmental niche, then the \emph{coupled phenotype-environment space} can be considered as an analogy of a meta-system. Here, by an environment we mean a set of conditions which may range from simple environmental variables, like temperature and humidity, to more complex holistic niche conditions. Since it encompasses the automaton, the coupled phenotype-environment space can operate as a first-order o-machine, in which the oracle provides input to the automaton and resolves computational undecidability. We assume that the meta-system would essentially be performing meta-simulation of the universal automaton, by observing and inverting the output of the universal automaton $U$ running on its own encoding $U[U]$. That is, the coupled phenotype-environment space may operate as the inverter $I[I]$. The detection of the paradox therefore occurs outside the bounds of the automaton.

Here we propose the question of delineating the nature of self-referential computational undecidability, generated by the $(\alpha - 1)$-order o-machine of the coupled phenotype-environment space. In other words, by what mechanism may a self-referential biological automaton generate a computational undecidability and by what form and channel is the corresponding oracle feedback transmitted?

In order to complete the expansion to the first-order system, the lower-order system (e.g., automaton) needs to receive a signal from the meta-level. While, as above, the form of such a signal is an open question, we can presume that the signal carries the information about the detected contradiction,  initiating  a \emph{generic} self-editing response \footnote{If the response were not generic, then some information about the contradiction would have to be known beforehand.}. The key element is an extension of the automaton's self-description with a new `axiom' so that the extended genotype better fits the niche. The resolution is implemented within the bounds of the extended, first-order system, which thus makes an evolutionary step by absorbing the innovation. This relationship recalls the tangled hierarchies of biological chemistry explored by Hofstadter \citeyearpar{hofstadter_go_1980} which are extended here to incorporate continual expansion in the $(\alpha - 1)$-order o-machine.
 
%Here we propose the second question. What is the nature of the information by which the biological automaton within the $(\alpha - 1)$-order o-machine augments feedback from the niche? In other words, what is the nature of the mutagenic program that incorporates feedback from the niche as an adaptive expansion of the genomic representation of the coupled phenotype-environment space?

A first-order system, of course, will have its own computational undecidability. For example,  questions about co-evolution of the biological automaton and its environmental niche may not be decidable within their first-order system, leading to some contradictions. However, a second-order system expanded with a more complex environmental context, i.e., second-order $o$-machine, will be able to resolve the ensuing contradictions, by providing  contextual co-evolutionary feedback, and generating further innovations at the level comprising co-evolving components.

To conclude, we highlight an insight from the exploration of `Life is Physics' by Goldenfeld and Woese \citeyearpar{goldenfeld_life_2011}: \begin{displayquote}
``These rules themselves need to evolve, but how? We need an additional set of rules describing the evolution of the original rules. But this upper level of rules itself needs to evolve. Thus, we end up with an infinite hierarchy, an inevitable reflection of the fact that the dynamic we are seeking is inherently self-referential.''\end{displayquote}
which also emphasises self-reference in biological computation. We argue that such self-reference inevitably generates undecidable dynamics, and hope that the questions raised by this Ansatz will help to progress this thread of enquiry.  

\section{Acknowledgements}
The Authors are grateful to Daniel Polani, Nihat Ay, Peter Stadler, Sheri Markose, Nathaniel Virgo, Felipe Abrah{\~a}o, Joseph Lizier, Michael Harr\'{e}, and Stuart Kauffman for helpful discussions of this cross-disciplinary topic. The authors were supported through the Australian Research Council grant DP200103005.

\linespread{1} % Set linespace
\bibliographystyle{apalike}
%\bibliography{220523_Ansatz_references.bib}

\appendix
\section*{Appendix}
\section{RNA-PDA for \texorpdfstring{$a^nb^n$}{a^(n),b^(n)}} 

To illustrate the use of an extensible memory encoding, the RNA-PDA we are constructing is designed to recognise input sequences of the form $a^nb^n$ where $n\geq 0$, in which the polymer must consist of an arbitrary number of $a$'s followed by an equal number of $b$'s. The automata must be able to encode the number of instances of $a$, and then compare this to the number of instances of $b$. Under this definition, we will accept an empty input, and reject a single $a$ or $b$. We will construct the RNA-PDA such that the automaton will halt at the end of the transition in which the end-of-input symbol $\nu$ is read. If the automaton is in an accept state at this point the input is considered to be accepted. The automaton does not halt immediately upon reaching an empty configuration, rather, explicit reject states may be reached from which no further input or stack symbol will result in a change of transition or stack operation.

$Q=\{q_0, q_1, q_2, q_3, q_4, q_5, q_6 \}$ where each $q_i$ is a unique sequence of the state polymer.\\$\Sigma = \{a, b, \nu\} \cup \{\epsilon\}$ where $a$ and $b$  are symbol polymers and $\nu$ is a special symbol polymer indicating the end of input.\\$\Gamma = \{a, \eta\}  \cup \{\epsilon \}$ where $a$ is a symbol polymer and $\eta$ is a special symbol polymer indicating the bottom of the stack.\\$Z_0 = \{\eta\}$.\\F = $\{ q_0, q_1\}$.

The transition function $\delta$ induces the following step-relations:

$\delta =
\begin{cases} 
\begin{aligned}
(q_0, aL, \epsilon S) &\vdash (q_0, L, aS) &&\text{where } 
    q_0 = \kappa (q_0, \kappa (a, \epsilon)), \\ 
    &&&\text{pushing a to the stack with } \lambda( \epsilon S, \kappa( a,  q_0)) \\
(q_0, bL, \eta S) &\vdash (q_2, L, \eta S) &&\text{where } 
    q_2 = \lambda (q_0, \lambda (b, \eta)), \\ 
	&&&\text{leaving the stack unchanged with } \kappa( \eta S, \kappa( b, q_0))	\\
(q_0, \nu L, aS) &\vdash (q_4,L, aS) &&\text{where } 
    q_4 = \lambda (q_0, \lambda (\nu, a)), \\ 
	&&&\text{leaving the stack unchanged with } \kappa( aS, \kappa( \nu, q_0))	\\
(q_0, bL, aS) &\vdash (q_1, L, \epsilon S) &&\text{where } 
    q_1 = \lambda (q_0, (\kappa(b, a)), \\ 
	&&&\text{popping from the stack with } \mu( aS, \kappa( b, q_0))	\\
(q_1, bL, aS) &\vdash (q_1, L, \epsilon S) &&\text{where } 
    q_1 = \kappa( q_1, (\kappa( b, a)), \\ 
	&&&\text{popping from the stack with } \mu(aS, \kappa( b, q_1))\\
(q_1, bL, \eta S) &\vdash (q_3, L, \eta S) &&\text{where } 
    q_3 = \lambda (q_1, \lambda (b, \eta)), \\ 
	&&&\text{leaving the stack unchanged with } \kappa( \eta S, \kappa( b, q_1))\\
(q_1, \nu L, aS) &\vdash (q_5, L, aS) &&\text{where } 
    q_5 = \lambda( q_1, \lambda( \nu, a)), \\ 
	&&&\text{leaving the stack unchanged with } \kappa( aS, \kappa( \nu, q_1))\\
(q_1, aL, \epsilon S) &\vdash (q_6, L, \epsilon S) &&\text{where } 
    q_6 = \lambda( q_1, \kappa( a, \epsilon)), \\ 
	&&&\text{leaving the stack unchanged with } \kappa( \epsilon S, \kappa( a, q_1))\\
\end{aligned}
\end{cases} 
$
\newpage
\begin{figure}[h!t]
\centering
    \includegraphics[]{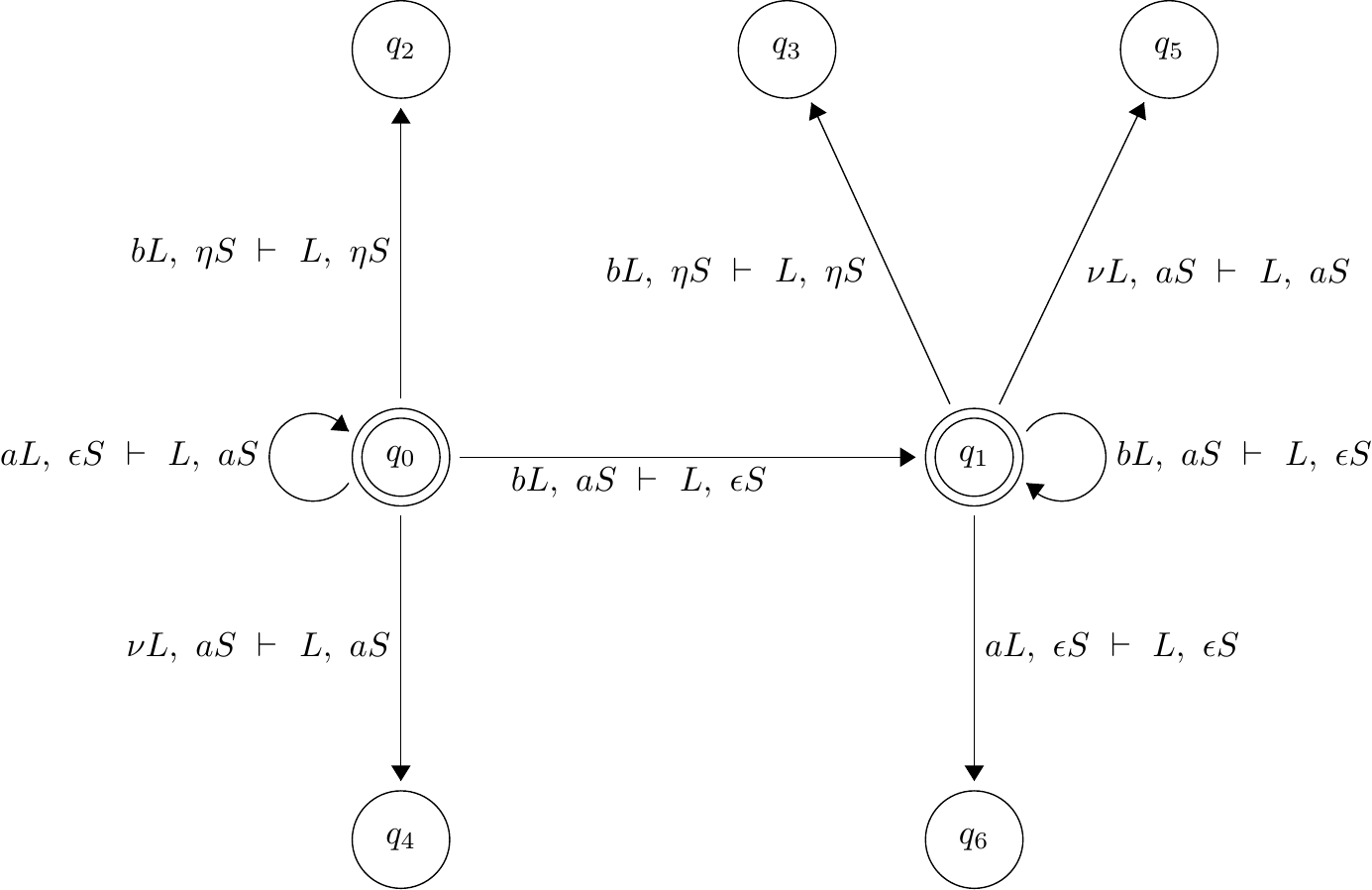}
    \caption{State diagram for the RNA-PDA.}
    \label {fig:RNAPDA}
\end{figure}
             
If we take as input \emph{$\nu$} (end of input symbol only): 
\begin{enumerate}
\item $(q_0, \nu L, \epsilon S)$ is empty, so no transition of state or stack occurs. 
\item At the exhaustion of input the state polymer has sequence $q_0$ so the automaton \textbf{accepts}.
\end{enumerate}

If we take as input \emph{aabb$\nu$}: 
\begin{enumerate}
\item $(q_0, aL, \epsilon S)$ results in the state polymer sequence remaining $q_0$ and the symbol polymer $a$ being placed on the stack, without reading the top of the stack. Stack is $a\eta$.
\item $(q_0, aL, \epsilon S)$ as for step 1. Stack is $aa\eta$.
\item $(q_0, bL, aS)$ results in ligation of the state polymer to sequence $q_1$ and the reading and popping of a $a$ from the stack. Stack is $a\eta$. 
\item $(q_1, bL, aS)$ results in the state polymer sequence remaining $q_1$ and the reading and popping of a $a$ from the stack. Stack in $\eta$.
\item $(q_1, \nu L, \epsilon S)$ is empty, so no transition of state or stack occurs.
\item At the exhaustion of input the state polymer has sequence $q_1$ so the automaton \textbf{accepts}.
\end{enumerate}

If we take as input \emph{abb$\nu$}: 
\begin{enumerate}
\item $(q_0, aL, \epsilon S)$ results in the state polymer sequence remaining $q_0$ and the symbol c being placed on the stack, without reading the top of the stack. Stack is $a\eta$.
\item $(q_0, bL, aS)$ results in ligation of the state polymer to $q_1$ and the reading and popping of a $a$ from the stack. Stack is $\eta$. 
\item $(q_1, bL, \eta S)$ results in ligation of the state polymer to $q_3$. No stack operation occurs.
\item $(q_3, \nu L, \epsilon S)$ is empty, so no transition of state or stack occurs.
\item At the exhaustion of input the state polymer has sequence $q_3$ so the automaton \textbf{rejects}.
\end{enumerate}

If we take as input \emph{aba$\nu$}: 
\begin{enumerate}
\item $(q_0, aL, \epsilon S)$ results in the state polymer sequence remaining $q_0$ and the symbol c being placed on the stack, without reading the top of the stack. Stack is $a\eta$.
\item $(q_0, bL, aS)$ results in the ligation of the state polymer to $q_1$ and the reading and popping of a $a$ from the stack. Stack is $\eta$. 
\item $(q_1, aL, \epsilon S)$ results in the ligation of the state polymer to $q_6$. No stack operation occurs. 
\item $(q_6, \nu L, \epsilon S)$ is empty, so no transition of state or stack occurs.
\item At the exhaustion of input the state polymer has sequence $q_6$ so the automaton \textbf{rejects}.
\end{enumerate}

\section{RNA-2PDA for \texorpdfstring{$a^{2^n}$}{a^(2^n)}}

The RNA-2PDA we are constructing will be to recognise the language $a^{2^n}$ consisting of sequences of $a$ in powers of 2. The RNA-2PDA must recursively divide the input word by 2, recognising when this results in an odd number to reject or in a single remaining symbol polymer.   Under this definition, we will reject the empty input $\nu$ and accept a single $a$. Any language that matches the form $a^{2^n}$ will result in an accept state.

$Q=\{q_0, q_1, q_2, q_3, q_4, q_5, q_6\}$ where each $q_i$ is a unique sequence of the state polymer.\\$\Sigma = \{a\}$ where $a$ is a symbol polymer and $\Sigma \in \Gamma$ as the input is initialised to the L tape.\\$\Gamma = \{a, b, \eta, \nu\}$ where $a$ and $b$ are symbol polymers and $\nu$ and $\eta$ are special symbol polymers indicating the bottom (3$'$-end) of the L and R stack respectively.\\$F = \{ q_6\}$.

The transition function $\delta$ is given by the following transitions:

$\delta =
\begin {cases}
\begin{aligned}
(q_0, \nu L,\eta R) &\vdash (q_1, \nu L, \eta R) &&\text {where } 
	q_1 = \lambda(q_0, \lambda(\nu L, \eta R)),\\
	&&&\text{leaving the stacks unchanged with } \kappa (\eta R, \kappa (\nu L, q_0)) 	\\
(q_0, \nu L, bR) &\vdash (q_2, \nu L, bR) &&\text {where } 
	q_2 = \lambda(q_0, \lambda(\nu L, bR)), \\
	&&&\text{leaving the stacks unchanged with } \kappa (\nu L, \kappa (bR, q_0)) 	\\
(q_0, aL, \epsilon R) &\vdash (q_3, \epsilon L, aR) &&\text {where } 
	q_3 = \lambda(q_0, \kappa(aL, \epsilon R)), \\
	&&&\text{popping L and pushing a to R with } \lambda (\epsilon R, \mu (aL, q_0)) 	\\
(q_2, \epsilon L, aR) &\vdash (q_2, aL, \epsilon R) &&\text {where } 
	q_2 = \kappa(q_2, \kappa(\epsilon L, aR)), \\
	&&& \text{popping R and pushing a to L with } \lambda (\epsilon L, \mu (aR, q_2)) 	\\
(q_2, \epsilon L, bR) &\vdash (q_2, L, \epsilon R) &&\text {where } 
	q_2 = \kappa(q_2, \kappa(\epsilon L, bR)), \\
	&&&\text{popping R with } \kappa (\epsilon L, \mu (bR, q_2)) 	\\
(q_2, aL, \eta R) &\vdash (q_0, aL, \eta R) &&\text {where } 
	q_0 = \mu(q_2, \lambda(aL, \eta R)), \\
	&&&\text{leaving the stacks unchanged with } \kappa (\epsilon L, \kappa (\eta R, q_2)) 	\\
(q_3, aL, aR) &\vdash (q_0, \epsilon L, bR) &&\text {where } 
	q_0 = \mu(q_3, \lambda(aL, aR)), \\
	&&&\text{popping L and pushing b to R with } \lambda (aR, \mu (aL, q_3)) 	\\
(q_3, \nu L, aR) &\vdash (q_4, \nu L, \epsilon R) &&\text {where } 
	q_4 = \lambda(q_3, \kappa(\nu L, aR)), \\
	&&& \text{popping R with } \kappa (\nu L, \mu (aR, q_3)) 	\\
(q_4, \nu L, bR) &\vdash (q_5, \nu L, bR) &&\text {where } 
	q_5 = \lambda(q_4, \lambda(\nu L, bR)), \\
	&&&\text{leaving the stacks unchanged with } \kappa (\nu L, \kappa (bR, q_4)) 	\\
(q_4, \nu L, \eta R) &\vdash (q_6, \nu L, \eta R) &&\text {where } 
	q_6 = \lambda(q_4, \lambda(\nu L, \eta R)), \\
	&&&\text{leaving the stacks unchanged with } \kappa (\nu L, \kappa (\eta R, q_4)) 	\\
\end{aligned}
\end{cases}
$
\begin{figure}[htp]
\centering
    \includegraphics[]{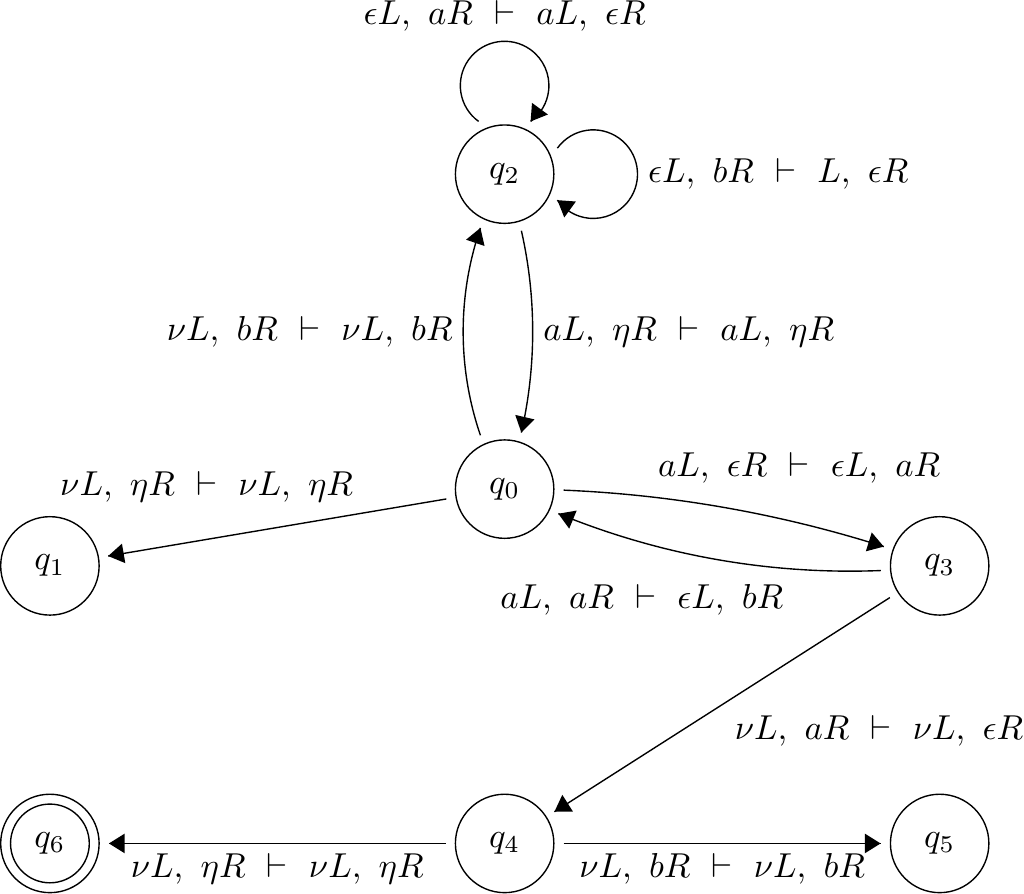}
    \caption{State diagram for the RNA-2PDA.}
    \label {fig:RNA-2PDA-TMsim}
\end{figure}

\newpage
If we take as input $a$: 
The stacks are initialised as L=$a \nu$ and R=$\eta$.
\begin{enumerate}
\item $(q_0, aL, \epsilon R)$ results in ligation of the state polymer to $q_3$ and $a$ being popped from L and pushed to R such that L=$\nu$, R=a$\eta$.
\item $(q_3, \nu L, aR)$ results in ligation of the state polymer to $q_4$ and $a$ being popped from R such that  L=$\nu$, R=$\eta$.
\item $(q_4, \nu L, \eta R)$ results in ligation of the state polymer to $q_6$ with no change of the stacks. 
\item There are no transitions possible from this configuration. The state polymer has sequence $q_6$ so the automaton \textbf{accepts.}
\end{enumerate}

If we take as input $a^2$:
The stacks are initialised as  L=$aa\nu$ and R=$\eta$.
\begin{enumerate}
\item $(q_0, aL, \epsilon R)$ results in ligation of the state polymer to $q_3$ and $a$ being popped from L and pushed to R such that L=$a\nu$, R=a$\eta$.
\item $(q_3, aL, aR)$ results in cleavage of the state polymer to $q_0$, $a$ being popped from L and $b$ being pushed to R such that  L=$\nu$ and R=$ba \eta$.
\item $(q_0, \nu L, bR)$ results in ligation of the state polymer to $q_2$ with no change to the stacks.
\item $(q_2, \epsilon L, bR)$ results in the state polymer sequence remaining $q_2$, $b$ being popped from R such that  L=$\nu$ and R=$a \eta$.
\item $(q_2, \epsilon L, aR)$ results in the state polymer sequence remaining $q_2$, with $a$ being popped from R and pushed to L such that  L=$a\nu$ and R=$\eta$.
\item $(q_2, aL, \eta R)$ results in cleavage of the stack polymer to $q_0$ with no change to the stacks.
\item The automaton now proceeds with input $a^1$, as above. As such, the automaton \textbf{accepts.}
\end{enumerate}

If we take as input $a^3$:
The stacks are initialised as  L=$aaa\nu$ and R=$\eta$.
\begin{enumerate}
\item $(q_0, aL, \epsilon R)$ results in ligation of the state polymer to $q_3$ and $a$ being popped from L and pushed to R such that L=$aa\nu$, R=a$\eta$.
\item $(q_3, aL, aR)$ results in cleavage of the state polymer to $q_0$, $a$ being popped from L and $b$ being pushed to R such that  L=$a\nu$ and R=$ba \eta$.
\item $(q_0, aL, \epsilon R)$ results in ligation of the state polymer to $q_3$ and $a$ being popped from L and pushed to R such that  L=$\nu$ and R=$aba\eta$.
\item $(q_3, \nu L, aR)$ results in ligation of the state polymer to $q_4$ and $a$ being popped from R such that  L=$\nu$, R=$ba\eta$.
\item $(q_4, \nu L, bR)$ results in ligation of the state polymer to $q_5$ with no change of stacks. 
\item There are no transitions possible from this configuration. The state polymer has sequence $q_5$ so the automaton \textbf{rejects.}
\end{enumerate}

For $n > 0$ , $a^{2^n}$ reduces to the computation of $a$.

\end{document}